\documentclass[twocolumn,amsmath,amssymb,superscriptaddress,aps]{revtex4}

\usepackage{amssymb}
\usepackage{graphicx}
\usepackage{dcolumn}
\usepackage{bm}
\newfont{\Fr}{eufm10}

\newcommand{\ba}{\begin{eqnarray}}
\newcommand{\ea}{\end{eqnarray}}
\def\be{\begin{equation}}
\def\ee{\end{equation}}

\begin{document}


\title{Anisotropic effects of background fields on Born-Infeld electromagnetic
waves}

\author{Mat\'\i as Aiello}
\email{aiello@iafe.uba.ar}
\thanks{ANPCyT Fellow}
\affiliation{Instituto de  Astronom\'\i a y F\'\i sica del Espacio,
Casilla de Correo 67, Sucursal 28, 1428 Buenos Aires, Argentina}
\affiliation{Departamento de F\'\i sica, Facultad de Ciencias Exactas y
Naturales, Universidad de Buenos Aires, Ciudad Universitaria, Pabell\'on
I, 1428 Buenos Aires, Argentina}
\author{Gabriel R. Bengochea}
\email{gabriel@iafe.uba.ar}
\thanks{CONICET Fellow}
\affiliation{Instituto de  Astronom\'\i a y F\'\i sica del
Espacio, Casilla de Correo 67, Sucursal 28, 1428 Buenos Aires,
Argentina} \affiliation{Departamento de F\'\i sica, Facultad de
Ciencias Exactas y Naturales, Universidad de Buenos Aires, Ciudad
Universitaria, Pabell\'on I, 1428 Buenos Aires, Argentina}
\author{Rafael Ferraro}
\email{ferraro@iafe.uba.ar}
\thanks{Member of Carrera del Investigador Cient\'{\i}fico (CONICET,
Argentina)} \affiliation{Instituto de  Astronom\'\i a y F\'\i sica
del Espacio, Casilla de Correo 67, Sucursal 28, 1428 Buenos Aires,
Argentina}
\affiliation{Departamento de F\'\i sica, Facultad de
Ciencias Exactas y Naturales, Universidad de Buenos Aires, Ciudad
Universitaria, Pabell\'on I, 1428 Buenos Aires, Argentina}


\begin{abstract}

We show exact solutions of the Born-Infeld theory for
electromagnetic plane waves propagating in the presence of static
background fields. The non-linear character of the Born-Infeld
equations generates an interaction between the background and the
wave that changes the speed of propagation and adds a longitudinal
component to the wave. As a consequence, in a magnetic background
the ray direction differs from the propagation direction --a
behavior resembling the one of a wave in an anisotropic medium--.
This feature could open up a way to experimental tests of the
Born-Infeld theory.

\end{abstract}

\pacs{Valid PACS appear here}
\keywords{Born-Infeld, waves,light propagation}
\maketitle

In 1934 Born and Infeld \cite{1born,2born} proposed a non-linear
electrodynamics with the aim of obtaining a finite value for the
self-energy of a point-like charge. The Born-Infeld Lagrangian
leads to field equations whose spherically symmetric static
solution yields a finite value $b$ for the electrostatic field at
the origin. The constant $b$ appears in the Born-Infeld Lagrangian
as a new universal constant. Following Einstein, Born and Infeld
considered the metric tensor $g_{\mu\nu}$ and the electromagnetic
field tensor $F_{\mu\nu}=\partial
_{\mu}A_{\nu}-\partial_{\nu}A_{\mu}$ as the symmetric and
anti-symmetric parts of a unique field $b\:g_{\mu\nu}+F_{\mu\nu}$.
Then they postulated the Lagrangian density
\begin{equation}
{\cal L}=-\frac{1}{4\pi}\bigg[\sqrt{|\det (b\:g_{\mu\nu}+F_{\mu\nu})|}-\sqrt{-\det
(b\:g_{\mu\nu})}\bigg]
\label{BIL}
\end{equation}
where the second term is chosen so that the Born-Infeld Lagrangian
tends to the Maxwell Lagrangian when $b\rightarrow \infty$. In
four dimensions, this Lagrangian results to be
\begin{equation}
{\cal
L}=\sqrt{-g}\frac{b^2}{4\:\pi}\bigg(1-\sqrt{1+\frac{2S}{b^2}-
\frac{P^2}{b^4}}\bigg)\label{BIL2}
\end{equation}
where $S$ and $P$ are the scalar and pseudoscalar field invariants
\begin{equation}
S=\frac{1}{4}F_{\mu\nu}F^{\mu\nu}=\frac{1}{2}(|{\bf B}|^2-|{\bf
E}|^2)
\end{equation}
\begin{equation}
P=\frac{1}{4}\:^{*}F_{\mu\nu}F^{\mu\nu}={\bf E}\cdot{\bf B}
\end{equation}
where the dual tensor is $^*\!F_{ij}= 1/2\ \varepsilon_{ijkl}\
F^{kl}$.

One of the typical features of non-linear electrodynamics is the
appearance of bi-refringence. However  the Born-Infeld Lagrangian
is usually mentioned as an exceptional Lagrangian because of the
properties of being the unique structural function which
\cite{4pleb}: 1- Assures that the theory has a single
characteristic surface equation (absence of bi-refringence); 2-
Fulfills the positive energy density and the non-space like energy
current character conditions. Due to these conditions, the
Lagrangian has time-like or null characteristic surfaces. The
Born-Infeld electrodynamical equations can be augmented to a
system of hyperbolic conservation laws with interesting properties
\cite{12brenier}.

It is a well established fact in non-linear electrodynamics that
the presence of background fields modifies the speed of
electromagnetic waves. This issue is studied in
\cite{4pleb,5pleb,6novello,7novello,8novello,9boi} by considering
the propagation of discontinuities. The result is that the phase
velocity is lower than $c$. Besides, the wave four-vector
direction can be described as a null geodesic of an effective
geometry that depends on the background field \cite{6novello}.
When there is no background field, the plane wave solution is the
same for the Maxwell and the Born-Infeld theory, as was pointed
out by Schr\"odinger \cite{3schro}. Some conditions for the
existence of global smooth spatially periodic planar solutions are
studied in \cite{12brenier}

In this work we find out exact solutions for Born-Infeld waves
propagating in the presence of a background field. As a still
unknown feature, we find that the presence of a magnetic
background modifies the ray direction, which does not result to be
coincident with the propagation direction (as it would happen in
an anisotropic medium).

The Born-Infeld field $F$ satisfies
\begin{eqnarray}
&&F_{\lambda\mu,\nu}+F_{\nu\lambda,\mu}+F_{\mu\nu,\lambda}=0\,
,\label {diffconindices}\\ &&(\sqrt{-g}\, {\cal
F}^{\mu\nu})_{,\nu}=0\, ,\label{difdualcurconindices}
\end{eqnarray}
where ${\cal F}_{\mu\nu}$ stand for the components of the 2-form
${\cal F}$ (antisymmetric 2-index covariant tensor) defined as
\begin{equation}\label{FBIcurs}
{\cal
F}=\frac{F-\frac{P}{b^2}\:^{*}F}{\sqrt{1+\frac{2\:S}{b^2}-\frac{P^2}{b^4}}}
\end{equation}
The equation (\ref{diffconindices}) is an identity coming from the
definition of $F_{\mu\nu}$, while (\ref{difdualcurconindices}) is
the Euler-Lagrange equation that results from varying the
Lagrangian (\ref{BIL2}).

We will do an extensive use of geometric language for benefiting
from some properties of the exterior derivative $d$:
$d(d...)\equiv 0$, $d\omega\wedge d\omega\equiv 0$ for any 1-form
$\omega$ (wedge product is the antisymmetrized tensor product
between $p$-forms). In this language it is $F=dA$, and the
equations (\ref{diffconindices}) read
\begin{eqnarray}
dF&=&0 \label {diff}
\\
d\:^{*}{\cal F}&=&0 \label{difdualcur}
\end{eqnarray}

We will use Cartesian coordinates in Minkowski space-time,
\begin{equation}\label{metric}
ds^2\ =\ dt^2\, -\, dx^2\, -\, dy^2\, -\, dz^2
\end{equation}

\begin{figure}[b]
\begin{center}
\includegraphics[width=5cm,angle=0]{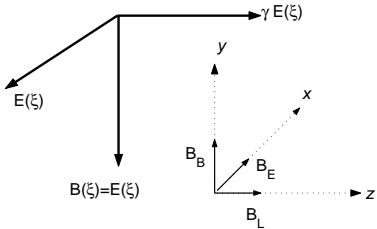}
\end{center}
\caption{Schematic picture of the wave and the background fields.}
\label{onda}
\end{figure}

We are looking for waves propagating in a uniform background
magnetic field. So we propose the solution
\begin{eqnarray}
\nonumber && F\ =\ E(\xi)\ d\xi\wedge dx\ +\ B_B\ dx\wedge dz\ -\
B_E\ dy\wedge dz \\ && -\ B_L\ dx\wedge dy \ +\ \gamma\ E(\xi)\
dt\wedge dz \label{solution1}
\end{eqnarray}
where
\begin{equation}
\xi\ =\ z - \beta\ t \label{variable}
\end{equation}
is the sole variable in the solution and $B_B$, $B_E$, $B_L$ are
the components of the uniform background magnetic field ${\bf{\cal
B}} = B_E\, \hat{x} + B_B\, \hat{y} + B_L\, \hat{z}$. The equation
(\ref{solution1}) means $F_{xz}=-E(\xi)+B_B$, $F_{tx}=-\beta\,
E(\xi)$, etc. So the electric field is $-E(\xi)\hat{x}+\gamma\,
E(\xi)\hat{z}$, and the magnetic field is
$-E(\xi)\hat{y}+{\bf{\cal B}}$. Figure \ref{onda} shows the
orientation of the wave and the background fields.
The terms having $E(\xi)$ compose the wave; $\xi$ is its phase.
$\beta\leq 1$ in (\ref{variable}) takes into account the fact that
waves might propagate inside the light cone. The solution
(\ref{solution1}) fulfills (\ref{diff}) for any function $E(\xi)$
(since $d\xi\wedge d\xi\equiv 0$, etc.).

The invariants $S$ and $P$ for the proposed solution are
\begin{equation}
2\, S\ =\ \left(1\, -\, \gamma ^2 - \beta^2\right)\, E(\xi)^2 +
{\cal B}^2 -2 E(\xi) B_B
\end{equation}
where ${\cal B}^2=B_L^2+B_E^2+B_B^2$, and
\begin{equation}
P\ =\ \left(-\beta B_E\ +\ \gamma B_L\right)\ E(\xi).
\end{equation}
In order to work out (\ref{difdualcur}), we will compute the
numerator in $^{*}{\cal F}$. According to (\ref{FBIcurs}) it is
\begin{eqnarray}
\nonumber&&^{*}\left(F\ -\ \frac{P}{b^2}\ ^{*}F\right)\ =\ ^{*}F\
+\ \frac{P}{b^2}\ F\\ \nonumber &&=\ E(\xi)\,  d(t-\beta\,
dz)\wedge dy - B_B\,  dt\wedge dy - B_E\,  dt\wedge dx \\
\nonumber && - B_L\, dt\wedge dz - \gamma\,  E(\xi)\,  dx\wedge dy
+\, b^{-2} \left(-\beta B_E + \gamma B_L\right)\  \\
\nonumber &&  E(\xi)  \big(E(\xi)\, d\xi\wedge dx + B_B\, dx\wedge
dz - B_E\, dy\wedge dz - \\ && B_L\, dx\wedge dy +\, \gamma \,
E(\xi)\, dt\wedge dz\big)
\end{eqnarray}
By replacing $dz$ with $d\xi\, +\, \beta\, dt$, the previous
result can be rewritten as
\begin{eqnarray}
&& \nonumber ^{*}\left(F\ -\ \frac{P}{b^2}\ ^{*}F\right)   \\
\nonumber &&=\ \left(-B_E\, -\, b^{-2}\, \beta\, B_B\, (-\beta
B_E\ +\ \gamma B_L)\ E(\xi)\right)\ dt\wedge dx \\ \nonumber &&+
\left((1-\beta^2) E(\xi) - B_B - b^{-2} \beta B_E (\beta B_E -
\gamma B_L) E(\xi)\right)\ \\ \nonumber && dt\wedge dy +\,
\left(-\gamma\, E(\xi)\, +\, b^{-2}\, B_L\, (\beta B_E\ -\ \gamma
B_L)\ E(\xi)\right)\ \\ && dx\wedge dy\ +\ d\xi\wedge
....\label{numerador1}
\end{eqnarray}
Since $d(d\xi\wedge ....)\equiv 0$, because $\xi$ is the only
variable in the field $F$, then the fulfillment of
(\ref{difdualcur}) exclusively depends on the behavior of the
three first terms in the former result. Taking into account that
the differentiation of these terms with respect to $\xi$ will
produce three independent components, then (\ref{difdualcur}) can
only be satisfied if $^{*}{\cal F}$ has components $tx$, $ty$ and
$xy$ equal to constants. Remarkably, the component $xy$ in
(\ref{numerador1}) is linear and homogeneous in $E(\xi)$, but this
feature is not shared with the denominator in (\ref{FBIcurs}). So,
in order to get $^{*}{\cal F}_{xy}= constant$, it should be
$E(\xi)=0$ or
\begin{equation}
\gamma\ =\ \frac{\frac{B_E B_L}{b^2}}{1+\frac{B_L^2}{b^2}}
\beta\label{gamma}
\end{equation}
This value for $\gamma$ can be replaced in the components
$^{*}{\cal F}_{tx}$ and $^{*}{\cal F}_{ty}$, which will turn to be
constants for $\beta$ equal to
\begin{equation}
\beta\ =\ \sqrt{\frac{1+\frac{B_L^2}{b^2}}{1+\frac{{\cal
B}^2}{b^2}}} \label{beta}
\end{equation}
In fact, the values of the examined components result to be
independent of the function $E(\xi)$, and (\ref{difdualcur}) is
accomplished:
\begin{equation}
^{*}{\cal F}_{tx}= -\frac{B_E}{\sqrt{1+\frac{{\cal B}^2}{b^2}}},\
\ ^{*}{\cal F}_{ty}=-\frac{B_B}{\sqrt{1+\frac{{\cal B}^2}{b^2}}},\
\ ^{*}{\cal F}_{xy}=0\label{ftxy}
\end{equation}
The obtained values for $\gamma$ and $\beta$ imply that
\begin{equation}
\sqrt{1+\frac{2S}{b^2}- \frac{P^2}{b^4}}\ =\ \sqrt{1 + \frac{{\cal
B}^2}{b^2}}-\frac{B_B\, E(\xi)}{b^2 \sqrt{1 + \frac{{\cal
B}^2}{b^2}}}
\end{equation}
The value $\beta < 1$ in (\ref{beta}) is the speed of propagation of a
Born-Infeld electromagnetic wave in the presence of a uniform
magnetic background. The constant $\gamma\neq 0$ in (\ref{gamma})
implies the existence of a non-zero electric longitudinal
component of the wave due to its interaction with the background.
Of course, these differences with the Maxwellian behavior
disappear in the limit $b \rightarrow \infty$.

Let us now consider the energy flux for a Born-Infeld wave in the
presence of a background field. The energy-momentum tensor in the
Born-Infeld theory is
\begin{eqnarray}
 && T^{\mu \nu}=\frac{2}{\sqrt{-g}}\, \frac{\delta{\cal
L}}{\delta g_{\mu\nu}} = \\ \nonumber && -\frac{1}{4 \pi} \bigg
[F^\mu\, _\rho\ {\cal F}^{\nu\, \rho}+ b^2 \: g^{\mu \nu}\
\bigg(1-\sqrt{1+\frac{2S}{b^2}- \frac{P^2}{b^4}}\bigg)\bigg]
\label{tem}
\end{eqnarray}
In particular the energy flux components are quite simple; for
instance
\begin{eqnarray}
\nonumber && T^{tz}=-\frac{1}{4\, \pi}\, F^t\, _\rho\, {\cal
F}^{z\rho}=-\frac{1}{4\, \pi}\, \left(F^t\, _x\, {\cal F}^{z
x}+F^t\, _y\, {\cal F}^{z y}\right)= \\ && \frac{F_{tx}F_{xz}\,
+\, F_{ty}F_{yz}}{\sqrt{1+\frac{2S}{b^2}- \frac{P^2}{b^4}}}
\end{eqnarray}
Remarkably, the terms proportional to $P$ in the numerator cancel
out. This is a foreseeable feature because $P$ is a pseudoscalar
while the energy flux $\pmb{\mathcal S}$ is a polar vector. Thus
the Poynting vector is
\begin{equation}
\pmb{{\cal S}}=\frac{1}{4\, \pi}\, \frac{{\bf E}\times{\bf
B}}{\sqrt{1+\frac{2S}{b^2}- \frac{P^2}{b^4}}}
\end{equation}
Then the energy flux vector associated with the solutions obtained
above are:
\begin{equation} {\cal S}_x\, =\, \frac{1}{4 \pi}\, \frac{\gamma
E(\xi)(E(\xi)-B_B)}{\sqrt{1+\frac{2\:S}{b^2}-\frac{P^2}{b^4}}}
\label{energiatx}\end{equation}
\begin{equation} {\cal S}_y\, =\, \frac{1}{4
\pi}\, \frac{\beta E(\xi) B_L + \gamma E(\xi)
B_E}{\sqrt{1+\frac{2\:S}{b^2}-\frac{P^2}{b^4}}} \label{energiaty}
\end{equation}
\begin{equation}
{\cal S}_z\, =\, \frac{1}{4 \pi}\, \frac{\beta
E(\xi)(E(\xi)-B_B)}{\sqrt{1+\frac{2\:S}{b^2}-\frac{P^2}{b^4}}}
\label{energiatz}
\end{equation}
As it is usual when $E(\xi)$ is a periodic function, we will
consider just the temporal averaging of the energy flux. Differing
from what happens in the Maxwell theory, the non-linear features of the
Born-Infeld theory will lead to non-null transversal components of
$<\pmb{\cal S}>$. This characteristic can be easily perceived in
$<{\cal S}_x>$. In fact ${\cal S}_x$ is an energy flux along the
wave polarization direction due to the existence of a longitudinal
electric field, which is a consequence of the interaction between
the wave and the background field. For a monochromatic wave,
$E(\xi)=E_0\, \cos(\beta^{-1}\omega\xi)$, the averaged flux at the
lower order in $b^{-2}$ is
\begin{equation}
<{\cal S}_x>\ =\ \frac{E_0^2\, B_L\, B_E}{8\, \pi\, b^2}\ +\ {\cal
O}(b^{-4})
\end{equation}
\begin{equation}
<{\cal S}_y>\ =\ \frac{E_0^2\, B_L\, B_B}{8\, \pi\, b^2}\ +\ {\cal
O}(b^{-4})
\end{equation}
\begin{equation}
<{\cal S}_z>\ =\ \frac{E_0^2}{8\, \pi}\left[1\ -\ \frac{4 B_B^2+2
B_E^2+B_L^2}{2\, b^2}\right]\ +\ {\cal O}(b^{-4})
\end{equation}
In this approximation, the transversal part of $<\pmb{\cal S}>$ is
parallel to the transversal background field. The angle $\alpha$
between $<{\pmb {\cal S}}>$ and the direction of propagation is
\begin{equation}
\tan\alpha\ =\ \frac{B_L\sqrt{B_E^2+B_B^2}}{b^2}\ +\ {\cal
O}(b^{-4})\label{alpha}
\end{equation}
In the same way we can calculate the energy density:
\begin{eqnarray} \nonumber && T^{00}=
\frac{|{\bf E}|^2\, +\, b^{-2}\, P^2}{4\, \pi\,
\sqrt{1+\frac{2\:S}{b^2}-\frac{P^2}{b^4}}}-\frac{b^2}{4\, \pi}
\bigg(1-\sqrt{1+\frac{2S}{b^2}- \frac{P^2}{b^4}}\bigg)
\\ && =\frac{b^2}{4\pi}\left[\frac{1 + b^{-2}|{\bf
B}|^2}{\sqrt{1+\frac{2\:S}{b^2}-\frac{P^2}{b^4}}}-1 \right]
\label{energiatt}
\end{eqnarray}
Therefore the averaged energy density for the studied solution is
\begin{eqnarray}\nonumber &&<{T}^{00}>=\frac{E_0^2}{8 \pi}\bigg[ 1-\frac{3 B_B^2+B_E^2+B_L^2}{2\, b^2}
\bigg] + \frac{{\cal B}^2}{8\, \pi} \bigg[1 -\frac{{\cal B}^2}{4
b^2} \bigg]\\ &&+\, {\cal O}(b^{-4}) \label{ptt}\end{eqnarray}
which is lower than the corresponding Maxwellian energy density.
We remark that, at order $b^{-2}$, the modulus of the energy
velocity $<{\pmb {\cal S}}><T^{00}>_{wave}^{^{-1}}$ does not
differ from the phase velocity $\beta$.

Now we will just display the solutions for the case of a constant
electric background field ${\pmb {\cal E}} =E_E \hat{x}+E_B
\hat{y}+E_L \hat{z}$. Following the same procedure we applied in
the previous case, the solution of (\ref{diff}) and
(\ref{difdualcur}) is:
\begin{eqnarray} \nonumber && F\ =\ E(\xi)\, d \xi\wedge dx\, +\, E_E\, dt\wedge
dx\, +\, E_B\, dt \wedge dy\\ && +\, E_L\, dt \wedge dz\, +\,
\gamma E(\xi)\, dt\wedge dz \label{felec} \end{eqnarray} with
\begin{equation}
\beta= \sqrt{1-\frac{E_B^2+E_E^2}{b^2}}, \:\:\:\: \gamma=\frac{E_L
E_E}{b^2 \sqrt{1-\frac{E_B^2+E_E^2}{b^2}}} \label{betaelec}
\end{equation}
Then
\begin{equation}
\sqrt{1+\frac{2S}{b^2}-\frac{P^2}{b^4}}=\sqrt{1-\frac{{\cal
E}^2}{b^2}}\, \left(1+\frac{E_E E(\xi)}{\beta \: b^2}\right)
\end{equation}
and the constant components of $^{*}{\cal F}$ are
\begin{eqnarray} \nonumber && ^{*}{\cal
F}_{tx}= -\sqrt{\frac{1-\frac{E_E^2+E_B^2}{b^2}}{1-\frac{{\cal
E}^2}{b^2}}} E_B, \:\:\:\: ^{*}{\cal
F}_{xy}=-\frac{E_L}{\sqrt{1-\frac{{\cal E}^2}{b^2}}},
\\ &&  ^{*}{\cal F}_{ty}=
\sqrt{\frac{1-\frac{E_E^2+E_B^2}{b^2}}{1-\frac{{\cal
E}^2}{b^2}}}E_E \label{ftxyel} \end{eqnarray} where ${\cal
E}^2=E_B^2+E_L^2+E_E^2$.
\smallskip
The components of Poynting vector ${\pmb{\cal S}}$ are
\begin{eqnarray}
\nonumber && {\cal S}_x=\frac{1}{4\pi}\frac{(E_L+\gamma
E(\xi))E(\xi)}{\sqrt{1+\frac{2S}{b^2}-\frac{P^2}{b^4}}},\:\:\:\:\:
\:\:\:\:\:\: {\cal S}_y=0,  \\ && {\cal S}_z=-\frac{1}{4 \pi}
\frac{(E_E-\beta
E(\xi))E(\xi)}{\sqrt{1+\frac{2S}{b^2}-\frac{P^2}{b^4}}}\end{eqnarray}
As a remarkable feature, the factor $E_L+\gamma E(\xi)$ in the
numerator of ${\cal S}_x$ is proportional to the square root in
the denominator; thus these factors cancel out and ${\cal S}_x$
results to be linear and homogeneous in $E(\xi)$. So there is a
main difference according to the Born-Infeld wave propagates in a
magnetic or an electric background field: in the second case the
direction of the averaged energy flux coincides with the
propagation direction:
\begin{eqnarray}
\nonumber && <{\cal S}_x>=0, \:\:\:\:\:\:\:\:\:\:\:\:\:\:  <{\cal
S}_y>=0,
\\ && <{\cal S}_z>=\frac{E_0^2}{8\pi}\, \left(1+\frac{2 E_E^2+ E_L^2}{2 b^2}\right)+{\cal
O}(b^{-4})
\end{eqnarray}
where $E_0$ is the wave amplitude.

The speeds of propagation (\ref{beta}) and (\ref{betaelec}) can be
compared with the results obtained in \cite{4pleb} and
\cite{6novello}. These papers study the propagation of
discontinuities in the presence of background fields in a general
non-linear theory. It is shown that the equation accomplished by
the wave four-vector can be understood as if rays propagate along
null geodesics of an effective metric. In the case of the
Born-Infeld electrodynamics the effective metric
$\bar{g}_{\mu\nu}$ is:
\begin{equation} \bar{g}_{\mu\nu}=(b^2+\frac{1}{2}F_{\rho \sigma} F^{\rho
\sigma}) g_{\mu\nu}+ F_{\mu\lambda} F^{\lambda}\:_{ \nu}
\label{geomef} \end{equation} being $g_{\mu\nu}$ the space-time
metric, and $F_{\mu \nu}$ is the background electromagnetic field
where the rays propagate. For a ray propagating along the $z$
direction it is
\begin{equation} d\bar{s}^2=\bar{g}_{00}dt^2+\bar{g}_{zz}dz^2=0 \Rightarrow
\frac{dz}{dt}=\sqrt{-\frac{\bar{g}_{00}}{\bar{g}_{zz}}}
\label{geoefe} \end{equation} When the effective metric
(\ref{geomef}) is evaluated for a magnetic background it results
\begin{equation} \bar{g}_{00}=\frac{1}{b^2+{\cal B}^2} \:\:\:\:\:\:\:\:\:\:\:\:\:
\bar{g}_{zz}=\frac{-1}{b^2+B^2_L}
\end{equation}
Instead,  for an electric background it is
\begin{equation} \bar{g}_{00}=\frac{1}{b^2} \:\:\:\:\:\:\:\:\:\:\:\:\:
\bar{g}_{zz}=\frac{-1}{b^2-E_B^2-E_L^2}
\end{equation}
Thus the speeds of propagation (\ref{beta}) and (\ref{betaelec})
are reobtained. This already known consequence of the Born-Infeld
theory on the wave propagation in background fields is here added
with the knowledge of the exact solutions. These solutions reveal
that the background fields not only affect the speed of
propagation of the Born-Infeld waves but also they can produce an
angle between the ray direction and the propagation direction.
This is the case for magnetic backgrounds. Of course, all these
effects would be very weak (if they exist), since up to now the
Maxwell equations properly describe all of the known classical
electromagnetic phenomena. Constant $b$ is the key for passing
from the Maxwell theory to the Born-Infeld theory. If the tiny
angle $\alpha$ (\ref{alpha}) were measured in the laboratory then
a way to experimental tests of the Born-Infeld electrodynamics
would be opened. The ray deviation is the consequence of the last
term in (\ref{solution1}), which is a longitudinal electric field
that results from the coupling with the background magnetic field.
The longitudinal electric field together with the speed of
propagation smaller than $c$ are the imprints of the non-linear
behavior. At the lowest order in $b^{-2}$ the longitudinal
component of the electric field is $b^{-2}B_E B_L E(\xi)$. A
similar longitudinal electric field appears in an electric
background as well. In this case, however, there is no
contribution to the averaged Poynting vector, so the ray does not
deviate.

\smallskip
M.A. and G.R.B. are supported by ANPCyT and CONICET graduate
scholarships respectively. This work was partially supported by
Universidad de Buenos Aires (Proy. UBACYT X103) and CONICET (PIP
6332).

\end{document}